\documentclass[aps,prc,preprint,showpacs,superscriptaddress,amsmath,amssymb,nofootinbib]{revtex4}

\usepackage{graphicx}
\usepackage{dcolumn}
\usepackage{bm}
\begin{document}


\title{Gamow-Teller strengths in $^{24}$Na using the $^{24}$Mg($t$,$^{3}$He) reaction at 115 AMeV.}

\author{M.E. Howard}
\affiliation{Joint Institute for Nuclear Astrophysics, Michigan State University, East Lansing, MI
48824, USA} \affiliation{Department of Physics, The Ohio State University, Columbus, OH 43210, USA}
\author{R.G.T. Zegers}
\email{zegers@nscl.msu.edu} \affiliation{National Superconducting Cyclotron Laboratory, Michigan
State University, East Lansing, MI 48824-1321, USA} \affiliation{Department of Physics and
Astronomy, Michigan State University, East Lansing, MI 48824, USA} \affiliation{Joint Institute for
Nuclear Astrophysics, Michigan State University, East Lansing, MI 48824, USA}
\author{Sam M. Austin}
\affiliation{National Superconducting Cyclotron Laboratory, Michigan State University, East Lansing, MI 48824-1321, USA}
\affiliation{Joint Institute for Nuclear Astrophysics, Michigan State University, East Lansing, MI 48824, USA}
\author{D. Bazin}
\affiliation{National Superconducting Cyclotron Laboratory, Michigan State University, East Lansing, MI 48824-1321, USA}
\author{B.A. Brown}
\affiliation{National Superconducting Cyclotron Laboratory, Michigan State University, East
Lansing, MI 48824-1321, USA} \affiliation{Department of Physics and Astronomy, Michigan State
University, East Lansing, MI 48824, USA} \affiliation{Joint Institute for Nuclear Astrophysics,
Michigan State University, East Lansing, MI 48824, USA}
\author{A.L. Cole}
\altaffiliation[Present address: ]{Department of Physics, Kalamazoo College, Kalamazoo, MI 49006-3295, USA}
\affiliation{National Superconducting Cyclotron Laboratory, Michigan State University, East Lansing, MI 48824-1321, USA}
\affiliation{Joint Institute for Nuclear Astrophysics, Michigan State University, East Lansing, MI 48824, USA}
\author{B. Davids}
\affiliation{TRIUMF, Vancouver, BC, V6T 2A3, Canada}
\author{M. Famiano}
\altaffiliation[Present address: ]{Physics Department, Western Michigan University, Kalamazoo, MI 49008-5252, USA}
\affiliation{National Superconducting Cyclotron Laboratory, Michigan State University, East Lansing, MI 48824-1321,
USA}
\author{Y. Fujita}
\affiliation{Department of Physics, Osaka University, Toyonaka, Osaka 560-0043, Japan}
\author{A. Gade}
\affiliation{National Superconducting Cyclotron Laboratory, Michigan State University, East
Lansing, MI 48824-1321, USA} \affiliation{Department of Physics and Astronomy, Michigan State
University, East Lansing, MI 48824, USA} \affiliation{Joint Institute for Nuclear Astrophysics,
Michigan State University, East Lansing, MI 48824, USA}
\author{D. Galaviz}
\altaffiliation[Present address: ]{Centro de Fisica Nuclear da Universidade de Lisboa, 1649-003, Lisboa, Portugal}
\affiliation{National Superconducting Cyclotron Laboratory, Michigan State University, East Lansing, MI 48824-1321,
USA}
\affiliation{Joint Institute for Nuclear Astrophysics, Michigan State University, East Lansing, MI 48824, USA}
\author{G.W. Hitt}
\affiliation{National Superconducting Cyclotron Laboratory, Michigan State
University, East Lansing, MI 48824-1321, USA} \affiliation{Department of Physics and
Astronomy, Michigan State University, East Lansing, MI 48824, USA}
\affiliation{Joint Institute for Nuclear Astrophysics, Michigan State University,
East Lansing, MI 48824, USA}
\author{M. Matos}
\altaffiliation[Present address: ]{Oak Ridge National Laboratory, Oak Ridge, TN 37831-6354, USA}
\affiliation{National Superconducting Cyclotron Laboratory, Michigan State University, East Lansing, MI 48824-1321,
USA}
\author{S.D. Reitzner}
\altaffiliation[Present address: ]{Physics Department, University of Guelph, Ontario N1G 2W1 Canada}
\affiliation{Department of Physics, The Ohio State University, Columbus, OH 43210, USA}
\author{C. Samanta}
\affiliation{Saha Institute of Nuclear Physics, 1/AF Bidhannagar, Kolkota 700064, India}
\affiliation{Physics Department, Virginia Commonwealth University, Richmond, VA 23284-2000, USA}
\affiliation{Physics Department, University of Richmond, Richmond, VA 23173, USA}
\author{L.J. Schradin}
\affiliation{Department of Physics, The Ohio State University, Columbus, OH 43210, USA}
\author{Y. Shimbara}
\altaffiliation[Present address: ]{Graduate School of Science and Technology, Niigata University, Niigata 950-2181, Japan}
\affiliation{National Superconducting Cyclotron Laboratory, Michigan State University, East Lansing, MI 48824-1321,
USA}
\author{E.E. Smith}
\affiliation{Joint Institute for Nuclear Astrophysics, Michigan State University, East Lansing, MI
48824, USA} \affiliation{Department of Physics, The Ohio State University, Columbus, OH 43210, USA}
\author{C. Simenel}
\affiliation{National Superconducting Cyclotron Laboratory, Michigan State University, East
Lansing, MI 48824-1321, USA}
\affiliation{CEA, Irfu, SPhN Centre de Saclay, F-91191 Gif-sur-Yvette, France}
\date{\today}%

\begin{abstract}
Gamow-Teller transitions from $^{24}$Mg to $^{24}$Na were studied via the ($t$,$^{3}$He) reaction at 115~AMeV using a secondary triton beam produced via fast fragmentation of 150~AMeV $^{16}$O ions. Compared to previous ($t$,$^{3}$He) experiments at this energy that employed a primary $\alpha$~beam, the secondary beam intensity is improved by about a factor of five. Despite the large emittance of the secondary beam, an excitation-energy resolution of $\sim 200$~keV is achieved. A good correspondence is found between the extracted Gamow-Teller strength distribution and those available from other charge-exchange probes. Theoretical calculations using the newly developed USDA and USDB $sd$-shell model interactions reproduce the data well.

\end{abstract}

\maketitle

\label{sec:intro}
Charge-exchange reactions have proven to be excellent tools for probing spin-isospin-flip excitations in nuclei \cite{HAR01}. In particular Gamow-Teller (GT) transitions, which are associated with spin-flip ($\Delta S=1$), isospin-flip ($\Delta T=1$) and zero units of angular-momentum transfer ($\Delta L=0$) can probe excitation-energy regions not accessible through $\beta$-decay experiments. The extracted GT strength distributions test nuclear-structure models, provide important input for simulations of stellar evolution and neutrino-induced nucleosynthesis, and can be used to constrain calculations of matrix elements for 2$\nu$ and neutrinoless double $\beta$ decay.

For charge-exchange reactions in the $\Delta T_{z}=+1$ direction ($\beta^{+}$ direction), a variety of probes are available of which the ($n,p$) \cite{JAC88,ALF86} and ($d$,$^{2}$He) \cite{RAK02,GRE04} reactions have been most widely employed to obtain information about GT strength distributions. It has been shown \cite{SHE99,DAI98,NAK00,COL06,ZEG06} that the ($t$,$^{3}$He) reaction at 115~AMeV is also an attractive probe. Good energy resolution ($\sim 200$ keV) can be achieved and experience with the ($^{3}$He,$t$) reaction at 140-150~AMeV \cite{FUJ96,FUJ07}, including a detailed study of the extraction of GT strength over a wide target-mass region \cite{ZEG07}, is of great benefit to the interpretation of ($t$,$^{3}$He) experiments.

The main challenge for the ($t$,$^{3}$He) experiments at intermediate beam energies is the use of a secondary triton beam, which results in a relatively low beam intensity and large emittance of the triton beam. Before the construction of the Coupled Cyclotron Facility (CCF) at the NSCL, a 140~AMeV primary $\alpha$ beam was used and triton beam intensities of $\sim 1\times10^{6}$ pps were achieved \cite{SHE99,NAK00,COL06,ZEG06}. After the coupling of the K500 and K1200 cyclotrons \cite{CCF}, a triton intensity of $\sim 5\times10^{6}$ was achieved by fast-fragmentation of a 150~AMeV primary $^{16}$O beam, as reported in Ref.~\cite{HIT06}. Here, we report on the first extraction of GT strengths with the new secondary triton beam from a measurement of the $^{24}$Mg($t$,$^{3}$He) reaction. The results are compared with those of $^{24}$Mg($d$,$^{2}$He) \cite{RAK02a} and $^{24}$Mg($^{3}$He,$t$) \cite{ZEG08a} (the latter by employing isospin symmetry), and shell-model calculations. Since the details of the production and rate-optimization studies for the triton beam from fast fragmentation of $^{16}$O have been discussed in Ref.~\cite{HIT06}, here the focus is on the reconstruction of the excitation-energy spectra, angular distributions and the extracted GT strength distribution.

The secondary triton beam of 115~AMeV was transported to a 9.86 mg/cm$^{2}$ thick, 99.92\% isotopically-enriched $^{24}$Mg target located at the entrance of the S800 spectrometer \cite{BAZ03}. In order to obtain high resolution ($t$,$^{3}$He) data, the beam lines and the spectrometer were operated in dispersion-matching mode, which limits the
momentum acceptance to $\pm 0.25$\%, corresponding to a 3~AMeV kinetic-energy spread of the triton beam. The transmission from the A1900 fragment separator \cite{MOR03} to the S800 target was $40-50$\%. This was lower than the expected value of about 80\%, which was traced back to small misalignments of certain beam-line elements. This has been resolved recently; the improved transmissions will lead to further increases in the triton beam intensity for future experiments.
The $^{3}$He particles produced in the $^{24}$Mg($t$,$^{3}$He) reaction were detected and identified in the focal plane detector system of the S800 \cite{YUR99}. Two cathode readout drift chambers determine the positions and angles in the dispersive and non-dispersive directions in the focal plane. A $5^{th}$-order transfer matrix \cite{BER93} was used to reconstruct the $^{3}$He momentum, the track angles in non-dispersive ($\theta_{\text{nd}}$) and dispersive ($\theta_{\text{d}}$) directions, and the non-dispersive hit-position at the target ($x_{\text{nd}}$).  The beam-spot size on target in the dispersive plane is about 5 cm and particles hit the target nearly parallel to the beam axis. The beam is focussed on the target in the non-dispersive plane. In contrast to the dispersive plane, the angular spread of the beam in this direction leads to an uncertainty in the scattering angle of the $^{3}$He particle. This uncertainty was reduced by slightly defocusing the beam and correcting $\theta_{\text{nd}}$ based on the correlation with $x_{\text{nd}}$. The resolutions in $\theta_{\text{nd}}$ and $\theta_{\text{d}}$ then become comparable, combining for a scattering-angle resolution of $0.5^{\circ}$.
\begin{figure}
\includegraphics[scale=0.9]{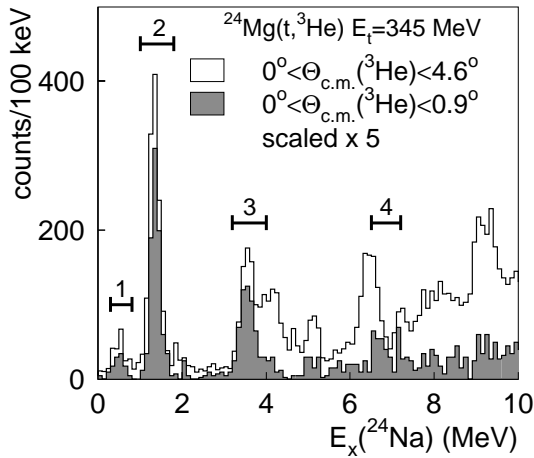}
\caption{\label{fig1}Excitation energy spectra in $^{24}$Na, measured via the $^{24}$Mg($t$,$^{3}$He) reaction at 115~AMeV, integrated over the full acceptance and at forward scattering angles, as labeled. The energy regions in which significant GT components in the spectra were measured are indicated and numbered 1-4.  }
\end{figure}
The tails of the 2-cm wide beam spot in the non-dispersive direction extended beyond the width of the $^{24}$Mg target (1.27 cm). As a result, a fraction of the beam hit the adhesive used to mount the target to the frame. The reconstruction of $x_{\text{nd}}$ (with a resolution of about 2~mm) is, therefore, also useful for removing the background events from the adhesive. It was hard to estimate how many $^{24}$Mg($t$,$^{3}$He) events were removed by this cut, leading to a large systematic error in the absolute cross sections.
The Gamow-Teller strengths were, therefore, normalized through comparison to existing data, as detailed below.

From the reconstructed angles and momentum, the $^{3}$He center-of-mass scattering angle $\theta_{c.m.}$($^{3}$He) and the excitation energy ($E_{x}$) of the $^{24}$Na were determined on an event-by-event basis. Due to the large size of the beam spot on the target, the acceptance of the spectrometer is not complete for angular ranges beyond $-3^{\circ}<\theta_{\text{d}}<3^{\circ}$ and $-3^{\circ}<\theta_{\text{nd}}<3^{\circ}$, which were, therefore, excluded from further analysis. The maximum $\theta_{c.m.}$($^{3}$He) covered was $4.6^{\circ}$.
\begin{figure}
\includegraphics[scale=1.0]{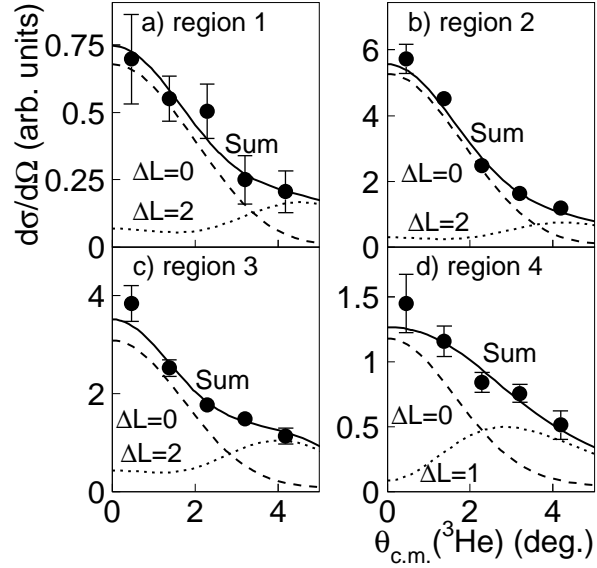}
\caption{\label{fig2} Angular distributions for each of the four excitation energy regions labeled in Fig.~\ref{fig1}.
The results of the MDA (solid line) and the constituent $\Delta L=0$ (long-dashed lines) and $\Delta L\ne0$ angular distributions (short-dashed lines) are superimposed (see text).}
\end{figure}
In Fig. \ref{fig1}, the excitation energy spectra over the full angular range and at forward angles
are displayed. Since GT transitions peak at forward scattering angles, unlike transitions associated with larger units of angular momentum transfer, comparison of these two spectra already gives an indication for the location of such states. Due to the kinematic correlation between $^{3}$He scattering angle and momentum and the finite angular resolution, the excitation energy resolution varied from 190 keV (FWHM) at the most forward scattering angles to 220 keV at backward angles.
\begin{figure}
\includegraphics[scale=1.0]{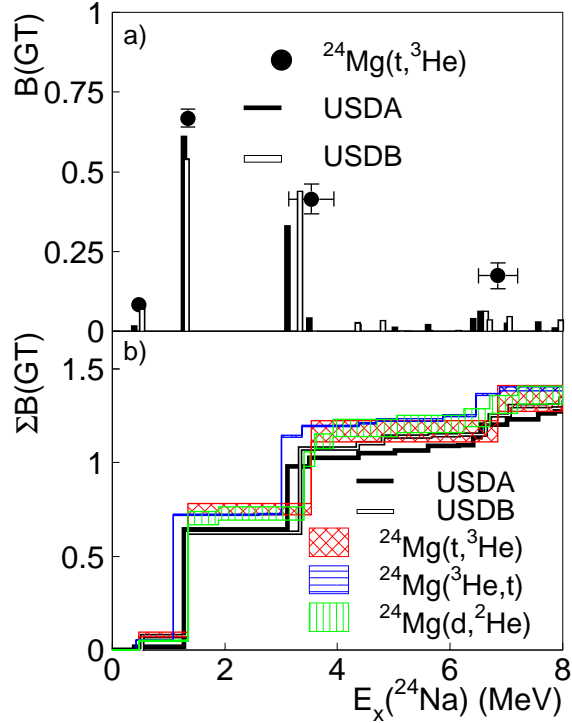}
\caption{\label{fig3}(color online) a) Extracted GT strength distribution from the present $^{24}$Mg($t$,$^{3}$He) experiment and the comparison with shell-model calculations using the USDA and USDB interactions. The experimental excitation energies of the lowest two GT transitions were fixed to the values known from Ref. \cite{ENSDF}. The two data points at higher energies are known to consist of two GT transitions (see text) and the horizontal bars indicate the energy intervals used in the analysis. b) Idem, but now for the cumulative strength. In addition, results from existing data using the $^{24}$Mg($d$,$^{2}$He) reaction \cite{RAK02a} and $^{24}$Mg($^{3}$He,$t$) reaction \cite{ZEG08a} are included. }
\end{figure}
The GT contribution to the spectrum is extracted from the angular distributions.
In four regions below 8~MeV, contributions from GT transitions are unambiguously identified and labeled 1-4 in Fig. \ref{fig1}. The angular distributions for these regions are displayed in Fig. \ref{fig2}. The extraction of the GT strength (B(GT)) relies on its proportionality to the GT cross section at zero momentum transfer ($q=0$) \cite{TAD87,ZEG07}. The GT cross section at $0^{\circ}$ are extracted from the data and is then extrapolated to $q=0$ based on the ratio $\frac{\sigma_{q=0}}{\sigma_{0^{\circ}}}$ calculated in Distorted Wave Born Approximation (DWBA).
No transitions with B(GT) known from $\beta$~decay are available to calibrate the proportionality. Instead, an indirect method was applied based on isospin symmetry of the strongest transition measured in the present work with that studied via the $^{24}$Mg($^{3}$He,$t$) reaction  \footnote{A detailed comparison between $^{24}$Mg($p,n$) \cite{AND91} and $^{24}$Mg($^{3}$He,$t$) experiments was performed in Ref.~\cite{ZEG08a}.} \cite{ZEG08a}. The B(GT) for the strongest GT transition in that reaction (at $E_{x}(^{24}\text{Al})=1.090(10)$~MeV) was deduced to be 0.668, using an empirical relationship for the unit cross section as a function of mass number \cite{ZEG07}. The error in that value is approximately 5\%, predominantly due to the uncertainty in the empirical relationship for the target-mass dependent unit cross section. As detailed in Ref.~\cite{ZEG06} for the case of the $^{26}$Mg($^{3}$He,$t$) reaction, the systematic errors in the extraction of GT strengths, mainly due to the effects of the tensor-$\tau$ component of the effective nucleon-nucleon interaction, are lowest for transitions with the largest B(GT). Hence, we performed the calibration with the strongest GT transition.

The four regions identified in Fig. \ref{fig1} are not of pure GT ($\Delta L=0$) nature. Besides the fact that the GT states are not completely isolated from neighboring states of different angular-momentum transfer due to the finite energy resolution, the GT transitions also contain minor quadrupole components because $0^{+}\rightarrow 1^{+}$ transitions can be due to the coupling of $\Delta L=2$ and $\Delta S=1$. Therefore, a simple multipole decomposition analysis (MDA) was performed for each of the four regions. The theoretical angular distributions used in the MDA were calculated in DWBA using the code \textsc{fold} \cite{FOLD} and were very similar to those performed for the $^{26}$Mg($t$,$^{3}$He) reaction in Ref.~\cite{ZEG06}. One-body transition densities (OBTDs) were calculated using the $sd$ shell-model interaction USDA \cite{BRO06} in an isospin-nonconserving proton-neutron formalism and the code \textsc{OXBASH} \cite{OXBA}. For the MDA, angular distributions were calculated using OBTDs of the states most closely matching the strength and excitation energies observed in the data. For dipole transitions, a pure $p_{3/2}$-$d_{5/2}$ transition was assumed.

The GT strength in region~1 (see Fig.~\ref{fig2}a) is due to the known $1^{+}$ state at 472 keV (\cite{ENSDF}), but in the present experiment this state cannot be separated from the nearby $2^{+}$ state at 563 keV. The MDA was performed, therefore, using angular distributions with $\Delta L=0$ and $\Delta L=2$, the latter representing both the quadrupole contribution to the GT excitation and the contribution from the nearby $2^{+}$ state. In region~2 (see Fig. \ref{fig2}b), a similar decomposition was performed; besides the known $1^{+}$ state at 1.346~MeV, non-separable states are present at 1.341~MeV ($2^{+}$) and 1.345~MeV (tentatively assigned as $3^{+}$ \cite{ENSDF}). The angular distributions for excitations of $2^{+}$ and $3^{+}$ states are quite similar \cite{ZEG06} and the choice of which angular distribution to use, besides the GT component, does not significantly affect the error in $\sigma_{0^{\circ}}$ beyond statistical uncertainties. In region~3 (see Fig. \ref{fig2}c), two $1^{+}$ states are known to exist (at 3.413~MeV and 3.589~MeV), which cannot be separated. In addition, several other weakly-excited states are present that are associated with various units of angular-momentum transfer, including possibly the tail of a dipole transition at $\sim$4~MeV. Fits with dipole or quadrupole contributions, in combination with the GT component (in Fig. \ref{fig2}c, the fit with a quadrupole contribution is shown) were performed. A difference of about $\pm 5$\% for $\sigma_{GT}(0^{\circ})$ was found and used as an estimate for the systematic error. For region~4, a relatively large contamination from the strong dipole transition at 6.5~MeV is to be expected and an MDA with GT and dipole contributions (see Fig. \ref{fig2}d) is probably more reasonable than with GT and quadrupole contributions. Nevertheless, based on the difference in the result for $\sigma_{GT}(0^{\circ})$ between the MDA using the different second multipole component, the systematic error is $\pm 10$\% for region 4.

After fixing the proportionality between B(GT) and $\sigma_{GT}(0^{\circ})$ for the strong GT transition at 1.346~MeV based on the strength extracted from the analog transition excited via $^{24}$Mg($^{3}$He,$t$), the B(GT)s in all four regions were deduced and are 0.08(1), 0.67(3), 0.41(5) and 0.17(4) for regions 1,2,3 and 4 respectively. The uncertainties do not include the $\pm 5$\% error due to the B(GT) normalization performed via the $^{24}$Mg($^{3}$He,$t$) data. These results are shown in Fig. \ref{fig3}a and compared with shell-model calculations using the USDA and USDB \cite{BRO06} interactions in $sd$ shell-model space. The theoretical results have been multiplied by 0.59 \cite{BRO88} to account for quenching of the GT strength due to configuration mixing with $2p-2h$ states and coupling to the $\Delta(1232)$-isobar nucleon-hole state. Both theoretical calculations reproduce well the experimental strength distribution. A convenient way to compare GT strength distributions is to plot cumulative sums, as is done in Fig. \ref{fig3}b. Besides the comparison of the present data with theory, the results from a $^{24}$Mg($d$,$^{2}$He) experiment \cite{RAK02a} and $^{24}$Mg($^{3}$He,$t$) experiment \cite{ZEG08a} are also included (the latter is based on the assumption of isospin symmetry). The results from the three data sets are plotted with errors due to statistical and fitting uncertainties only.
The energy resolution of the $^{24}$Mg($d$,$^{2}$He) experiment is 145 keV, slightly better than the 190 keV in the present experiment and much better than what is achievable using the ($n$,$p$) reaction ($\sim 1$ MeV). Combined with the smaller statistical uncertainties, some very weakly excited states seen in the ($d$,$^{2}$He) experiment are not separated in the present data. Nevertheless, the overall good agreement between the two data sets demonstrates that ($t$,$^{3}$He) reaction studies using a secondary beam of tritons produced from a $^{16}$O primary beam are appropriate for extracting GT strength distributions. The $^{24}$Mg($^{3}$He,$t$)  experiment had very high resolution (35 keV) and very small statistical uncertainty. Hence, the level of detail extracted is highest. The locations of the individual levels measured in the $\Delta T_{z}=-1$ channel are slightly shifted from those measured in the $\Delta T_{z}=+1$ channel because of Coulomb effects.

In summary, Gamow-Teller strengths have been extracted via the  $^{24}$Mg($t$,$^{3}$He) reaction using a 115~AMeV secondary triton beam produced from a primary $^{16}$O beam. Through comparison with existing data from $^{24}$Mg($d$,$^{2}$He) and $^{24}$Mg($^{3}$He,$t$) experiments employing stable beams, it was shown that, in spite of the large beam emittance of the secondary beam, detailed measurements of GT strengths and tests of shell-model calculations are feasible. Problems with the alignment of the beam lines to the S800 and the size of the target, which reduced the triton beam intensity and made it hard to extract absolute cross sections, were identified and will be corrected in future experiments.

We thank the NSCL cyclotron and A1900 staff for their efforts during the experiment. This work was supported by the US-NSF (PHY0216783 (JINA), PHY0606007, PHY0140255 and PHY0758099.)

\bibliography{prc}

\end{document}